\documentclass[reprint,amsmath,amssymb,prb]{revtex4-1}
\usepackage[pdftex]{graphicx}
\pdfoutput=1
\begin{document}

\preprint{}

\title{Extension of structure integration to magnetic system}


\author{Kazuhito Takeuchi}
\author{Takashi Ishikawa}
\author{Ryohei Tanaka}
\author{Koretaka Yuge}
%\thanks{}
\affiliation{Department of Materials Science and Engineering, Kyoto University, Sakyo, Kyoto 606-8501, Japan}


\date{\today}

\begin{abstract}
We extend our previously developed method, "structure integration", to evaluate free energy directly for magnetic large systems on given lattice.
The present method express the density of states(DOS) as the parameters independent of system but depend on lattice, and replace the DOS unknown {\it a priori} as that known {\it a priori}.
Through two-dimensional square lattice Ising model, we find that the present method can evaluate magnetic free energy efficiently and accurately above critical temperature without iterative method like Monte Carlo simulation.

\end{abstract}

%\keywords{}

\maketitle

\section{introduction}
\label{sec:introduction}
Helmholtz free energy, $F$, is the most important and essential property in thermodynamics.
In statistical mechanics, however, estimating Helmholtz free energy is a challenging problem, because $F$ is not statistically averaged value that is easily obtained by Metropolis algorithm\cite{:/content/aip/journal/jcp/21/6/10.1063/1.1699114}. $F$ is derived from partition function, $Z$, that is defined by:
\begin{equation}
\label{eq:partition_function}
	Z = \sum_E W(E)\exp \left( -\frac{E}{k_{\rm B}T} \right),
\end{equation}
where $W$ is the number of states, $k_{\rm B}$ is boltzmann constant and $E$ is total energy.
Here $W(E)$ is needed to obtain $Z$, however $W(E)$ is not known {\it a priori}.
The most simple approach to obtain $W(E)$ is to search phase space and caluculate $E$ for all possible states.
This approach requires enormous computational cost, therefore this approach is limited for small system because of practical difficulty.
Microscopic structure of substitutional crystalline solids is often represented by Ising Hamiltonian, where phase space can be splited into momentum space and configuration space\cite{solid-state}.
In order to estimate $Z$ on Ising Hamiltonian for large system, two main methods exist; (i) analytical methods and (ii) numerical methods.
In (i),
for example, transfer-matrix method\cite{PhysRev.65.117} is one of the most successful method to give exact solution of free energy.
However, since such analytical methods are limited to the system whose Hamiltonian is quite simple, analytical methods cannot be applied to the system whose Hamiltonian has multibody interaction terms.
This difficulty can be avoided using (ii) such as multicanonical ensemble\cite{BERG1991249, PhysRevLett.68.9, PhysRevLett.69.2292}, entropic sampling\cite{PhysRevLett.71.211}, Wang-Landau method\cite{PhysRevLett.86.2050, PhysRevE.64.056101} and so on.
Although these method effectively sample microscopic states over vast configuration space, these methods cannot avoid the problem with computational cost that increase in system size.
This is because the number of points in cofiguration space exponentially increase with increase of system size(e.g., in $N$ spin Ising model, the number of points in configuration space is $2^N$).
In addition, as the number of interactions increase, problems become too intractable.
As shown in (i) and (ii), estimating $Z$ suffers from increasing both $N$ and the number of interactions.

We have recently proposed a new approach, "structure integration(SI)\cite{0953-8984-27-38-385201}", overcoming above difficulties and enabling direct evaluation of configurational free energy for large binary alloy system.
SI gives universal and analytical representation of $W$ via so-called "correlation functions\cite{SDG}" which does not depend on both constituent elements or system size.
Since SI is based on information about density of microscopic states on configuration space established from crystal lattice\cite{Yuge_sro}, SI can be applied to any lattice system, e.g., fcc, bcc, hcp and square lattice with any number of multibody interactions.
In our previous study\cite{0953-8984-27-38-385201}, we applied SI to equiatomic Cu-Au alloy, that show first order phase transition from $L1_0$ to disorder states above critical temperature, $T_{\rm c}$, through first-principles-based simulation. Comparing with free energy using SI and exact method in small system, we found that SI could evaluate configurational free energy accurately in disorder states above $T_{\rm c}$.
This good agreement for disorder states implies that our analytical representationf of $W$ can describe the major part of true $W$.
However, since SI have been established based on "constant" composition, analytical representation of $W$ could not be applied for the system where composition can vary.

In present study, we give an analytical representation of $W$ for any composition and extend SI to magnetic system.
This extension enable us to estimate $Z$ in large spin system for multibody interactions.
In order to confirm validity and applicability, we apply the present method to a model system, two-dimensional(2D) square lattice Ising model that has been generally used as a benchmark for new simulation, because spin system cannnot be described by equicomponent method.

\section{methodology}
\label{sec:methodology}
In this section, first, we give a breif explanation of SI and generalized Ising model.
Second, the derivation of extension of SI to magnetic system is shown.
Finally, we show how the extended SI is applied to a simple model system.

	\subsection{Structure integration approach}
In order to describe thermodynamic property via microscopic states confined to atomic arrangements on a given lattice, generalized Ising model for multibody interactions\cite{SDG} has been proposed and widely used in alloy studies.
Then, configurational property, e.g., $E$, is competely represented via correlation functions, $\xi_k$,
\begin{equation}
\label{eq:CE}
	E = \sum_k V_k \xi_k,
\end{equation}
where $k$ and $V_k$ denote the figure of cluster(e.g., point, pair, triangle and tetora) and effective cluster interaction(ECI).
$\xi$s are product of complete basis function that are constructed by Gram-Schmidt orthonormalization.
Especially in binary alloy system, the basis functions are $\{ 1,\theta \}$ where $\theta$ is a spin variable that has $+1$ or $-1$.
Therefore in binary case, $\xi_k$ are expressed by the average for spin products in cluster $k$.
From the completeness and orthogonality of $\xi$s, $V_i$ is described by $\langle E|\xi_i \rangle$, where $\langle | \rangle$ denotes inner product.
This means that the configurational property can be split into the term depend only on lattice and on others.
In our previous study\cite{Yuge_sro}, we confirmed that the distribution of any $\xi_k$ numerically obey to normal distribution function(NDF) at large system size, $P_k(\xi_k)$, whose average, $\mu_k$, and standard deviation, $\sigma_k$, are as follows\cite{PhysRevB.42.9622}:
\begin{equation}
\label{eq:average}
	\mu_k = (2x-1)^k,
\end{equation}
\begin{equation}
\label{eq:standard_deviation}
	\sigma_k = \frac{1}{\sqrt{ND_k}} \;\;\; (x=0.5).
\end{equation}
\begin{figure}
	\includegraphics[clip, width=\columnwidth]{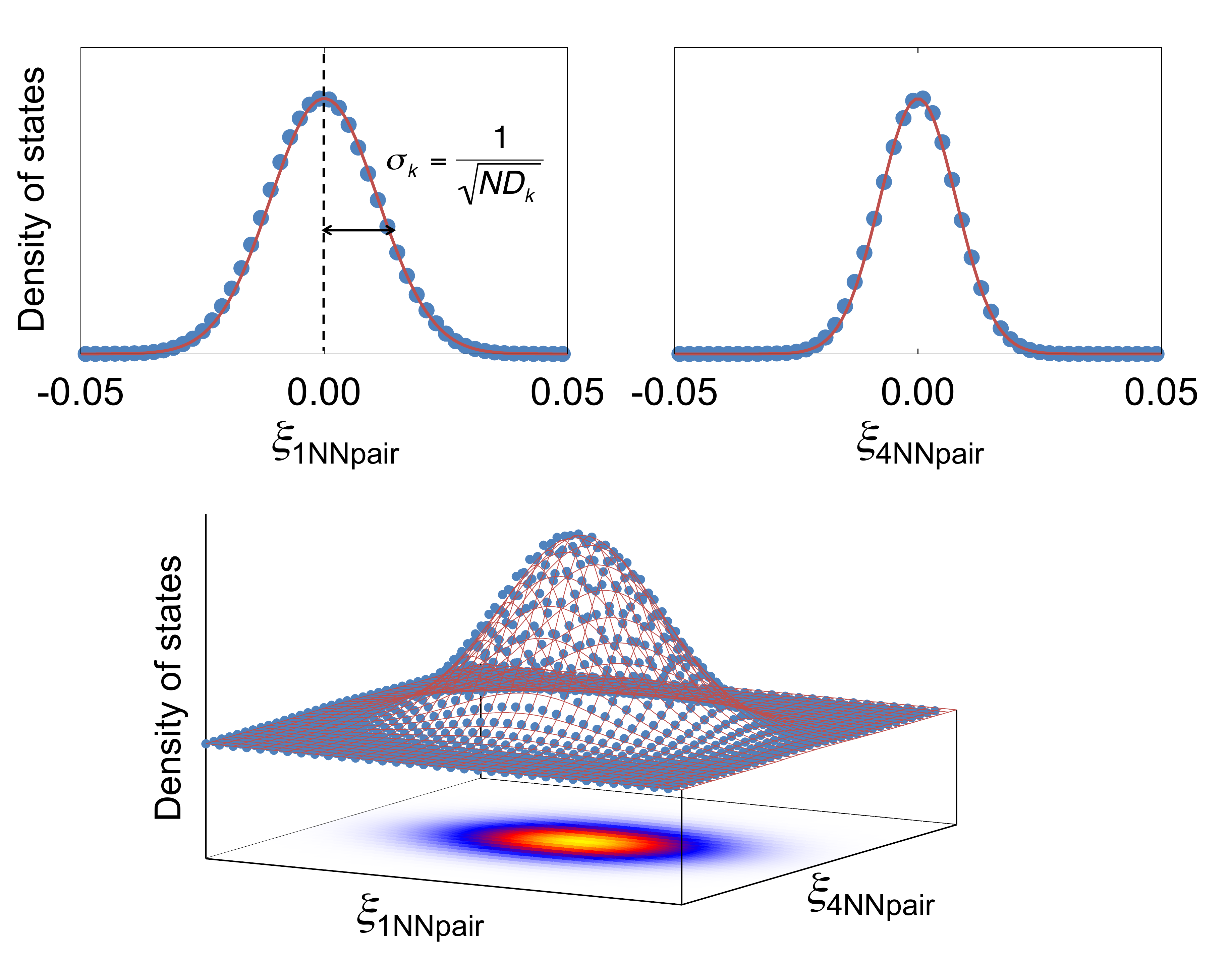}%
	\caption{\label{fig:dos_xi}
	Upper part: Density of states in terms of correlation function. Points denote the normalized number of states from MC simulation in ${\rm A}_{2048}{\rm B}_{2048}$ binary alloy system on 2D square lattice. Solid curves are normal distribution function whose average and standard deviation are derived from lattice.
	Lower part: Density of states in SI for $E = \sum_k V_k \xi_k$ where $k$ means the first and fourth nearest neighbor pair(1NNpair and 4NNpair) cluster.
	}
\end{figure}
Here $x$ is composition, $N$ is the number of sites and $D_k$ is the number of $k$ clusters per site.
In upper part of Fig.~\ref{fig:dos_xi}, we compare the distribution from Monte Carlo(MC) simulation and the DOS known {\it a priori}.
This good agreement is reliable in disorder states.
We confirmed that any single variable $P_k(\xi_k)$ correspond to marginal distribution of the DOS in terms of the set of $\xi$s, $\vec{\xi}$.\cite{Yuge_sro, 0953-8984-27-38-385201}
In SI, we approximate that any correlation coefficient between different $\xi$s are zero at thermodynamic limit, i.e.,\cite{Yuge_sro}
\begin{equation}
\label{eq:decomposition}
	P(\vec{\xi}) \simeq \prod_{k} P_k(\xi_k).
\end{equation}
From Eq.~(\ref{eq:partition_function}), (\ref{eq:CE}) and (\ref{eq:decomposition}), $Z$ is rewritten as\cite{0953-8984-27-38-385201}
\begin{equation}
\label{eq:SI}
	Z \simeq \prod_{k} \int P_k(\xi_k) \exp \left( -\frac{V_k\xi_k}{k_{\rm B}T} \right) d\xi_k.
\end{equation}
We call Eq.~(\ref{eq:SI}) "structure integration" because the integration variable means the basis of configuration.
Although $\xi_k$ is discrete variable, in Eq.~(\ref{eq:SI}) we employ integral instead of sum, because expressing the DOS as Eq.~(\ref{eq:decomposition}) is allowed at thermodynamic limit.
Eq.~(\ref{eq:SI}) can be applied only to equicomponent alloy, because Eq.~(\ref{eq:standard_deviation}) holds only for equiatomic composition.
Therefore, SI cannot be applied to magnetic system.

\subsection{Present extension of structure integration}
Here we extend Eq.~(\ref{eq:SI}) to any composition.
When the changes in composition is included to $Z$, $P_k(\xi_k)$ and intervals of integration should be rewritten to be in terms of $x$.
The characteristics of $P_k(\xi_k)$ are determined by $\mu_k$ and $\sigma_k$.

For simplicity and correspondence to Ising model, without the lack of generality, hereinafter we consider $E=V\xi$ where $\xi = \sum_{\langle i,j \rangle }\theta_i \theta_j$ and $\langle i,j \rangle$ denotes that a pair of $\theta_i$ and $\theta_j$ is nearest neighbor, i.e., $E$ is expressed by single correlation function.
In Ising model, $Z$ is derived from the sum of boltzmann factor, $\exp \left( -V\xi/k_{\rm B}T \right) $, over all possible configuration.
Meanwhile in SI, since configuration space is expressd as $(x, \xi)$, the DOS should be expressed in terms of $x$ and $\xi$; $P(x, \xi)$.
From previous study\cite{0953-8984-27-38-385201}, we know that $P(0.5, \xi)$ is described by NDF of which average and standard deviation are Eq.~(\ref{eq:average}) and (\ref{eq:standard_deviation}), and we also have shown that $P(x, \xi)$ where $x \neq 0.5$ can also be described by NDF in large system size.

\begin{figure}
	\includegraphics[clip, width=0.8\columnwidth]{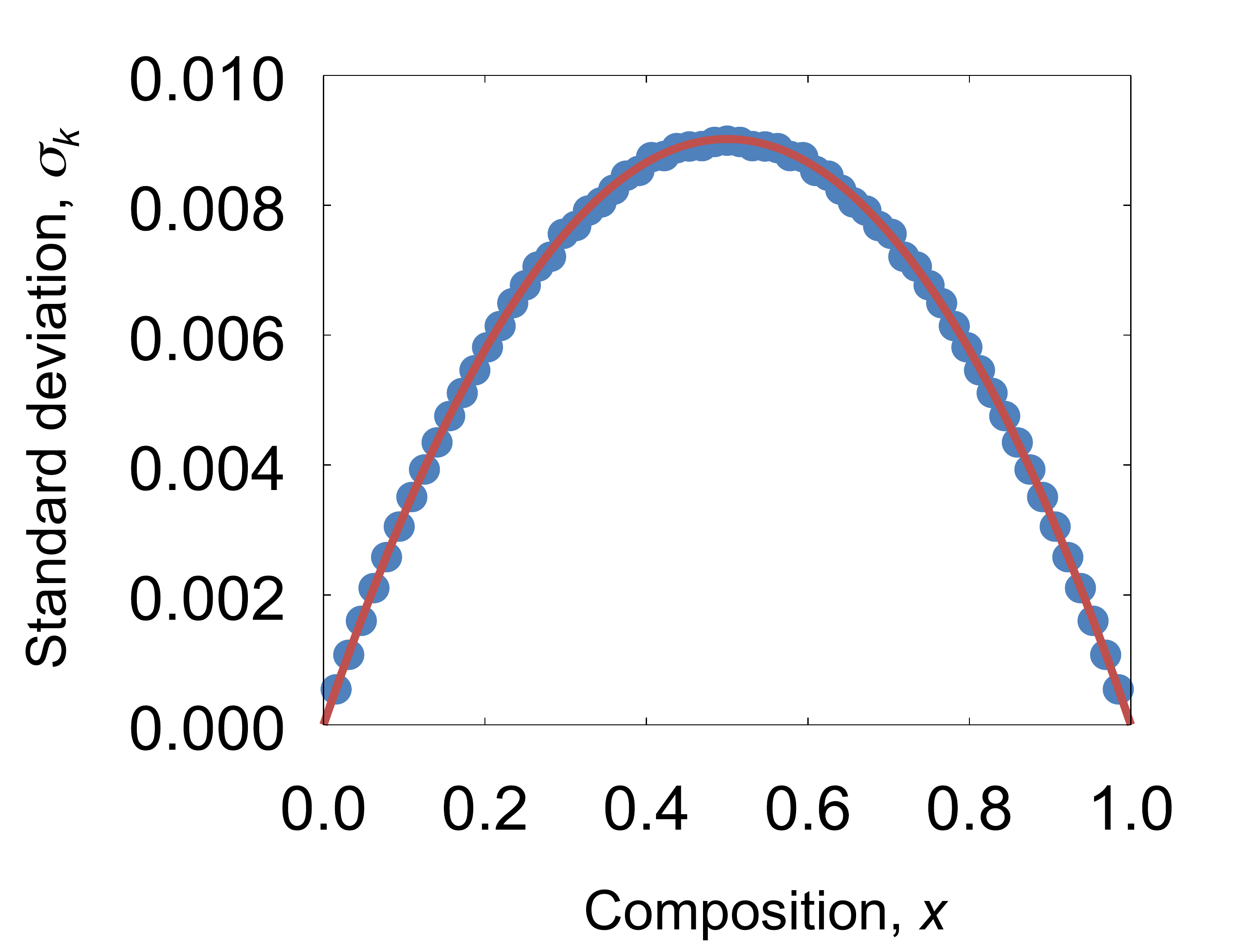}%
	\caption{\label{fig:var_vs_comp}
	Standard deviation as a function of composition in $A_xB_{1-x}$ binary alloy system on fcc lattice when $8 \times 8 \times 8$ supercell. Points and solid curve denote the result of MC simulation and fitted quadratic function(Eq.~(\ref{eq:var_vs_comp})).
	 }
\end{figure}

Since $\mu_k$ in Eq.~(\ref{eq:average}) is already described by $x$, $\sigma_k$ should be described in terms of $x$.
Through MC simulation for $N=2048$ atoms on fcc lattice with $A_xB_{1-x}$ binary system, $\sigma_k (x)$ for each $x$ is shown in Fig.~\ref{fig:var_vs_comp}.
In Fig.~\ref{fig:var_vs_comp}, we find that the points denote the standard deviations are well described as:
\begin{equation}
\label{eq:var_vs_comp}
	\sigma_k(x) \simeq \frac{1-4(0.5-x)^2}{\sqrt{ND_k}}.
\end{equation}
Eq.~(\ref{eq:var_vs_comp}) satisfies the condisitions that $\sigma_k(0)=0$, $\sigma_k(0.5)=\sqrt{ND_k}$ and $\sigma_k(1)=0$.
We also confirm that Eq.~(\ref{eq:var_vs_comp}) is satisfied on 2D square lattice.
The form of $P(x, \xi)$ for each $x$ is shown, and furthermore we should consider the total number of states for each $x$.
The total number of states is $_NC_{Nx}$, and this should be described by the gamma function as $_NC_{Nx} = \Gamma(N+1) / \Gamma(Nx+1)\Gamma(N(1-x)+1)$ in the SI formalism in order to employ integration.
Using Eq.~(\ref{eq:var_vs_comp}) and the gamma function, $P(x, \xi)$ is described as:
\begin{eqnarray}
\label{eq:DOS}
	P(x, \xi) &\simeq& \frac{1}{\sqrt{2\pi \sigma(x)^2}}
				\frac{\Gamma(N+1)}{\Gamma(Nx+1)\Gamma(N(1-x)+1)}  \nonumber \\
				&& \times \exp \left(-\frac{(\xi - \langle \xi \rangle)^2}{2 \sigma(x)^2}  \right).
\end{eqnarray}
Then $Z$ is rewritten as:
\begin{equation}
\label{eq:SI_mod}
	Z \simeq \int_x \int_\xi P(x, \xi) \exp \left( -\frac{V\xi}{k_{\rm B}T} \right) d\xi dx.
\end{equation}
An advantage of this new expression of $Z$ is that the computational cost does not depend on $N$, because $x$ and $\xi$ are intensive variables. In contrast, additivity of $F$ is explicitly considered in $P(x, \xi)$ and $V$.
Since $Z$ is analytically rewritten, our method does not need iterative method like MC simulation.
When the number of $V_k$ is more than one, Eq.~(\ref{eq:SI_mod}) becomes:
\begin{equation}
	Z \simeq \int_x \prod_k \int_{\xi_k} P_k(x, \xi_k) \exp \left( -\frac{V_k\xi_k}{k_{\rm B}T} \right) d\xi_k dx. \nonumber
\end{equation}
Here approximately equal is caused by Eq.~(\ref{eq:decomposition}).
As above, we emphasize that our method can be applied to any lattice and any number of interactions.
Note that SI cannot be applied if any correlation coefficient between different $\xi$s does not approach to zero at thermodynamic limit.
We confirmed that when
constituent lattice points on cluster $k$ are all included in those on cluster $k'$, the correlation coefficient between $k$ and $k'$ does not approach to zero at thermodynamic limit.
In our previous study\cite{0953-8984-27-38-385201}, we showed that this difficulty can be overcomed using a new basis set that makes any non zero correlation coefficients between different $\xi$s zero.

\section{\label{sec:level3} result and discussion}
\begin{figure}
	\includegraphics[clip, width=0.8\columnwidth]{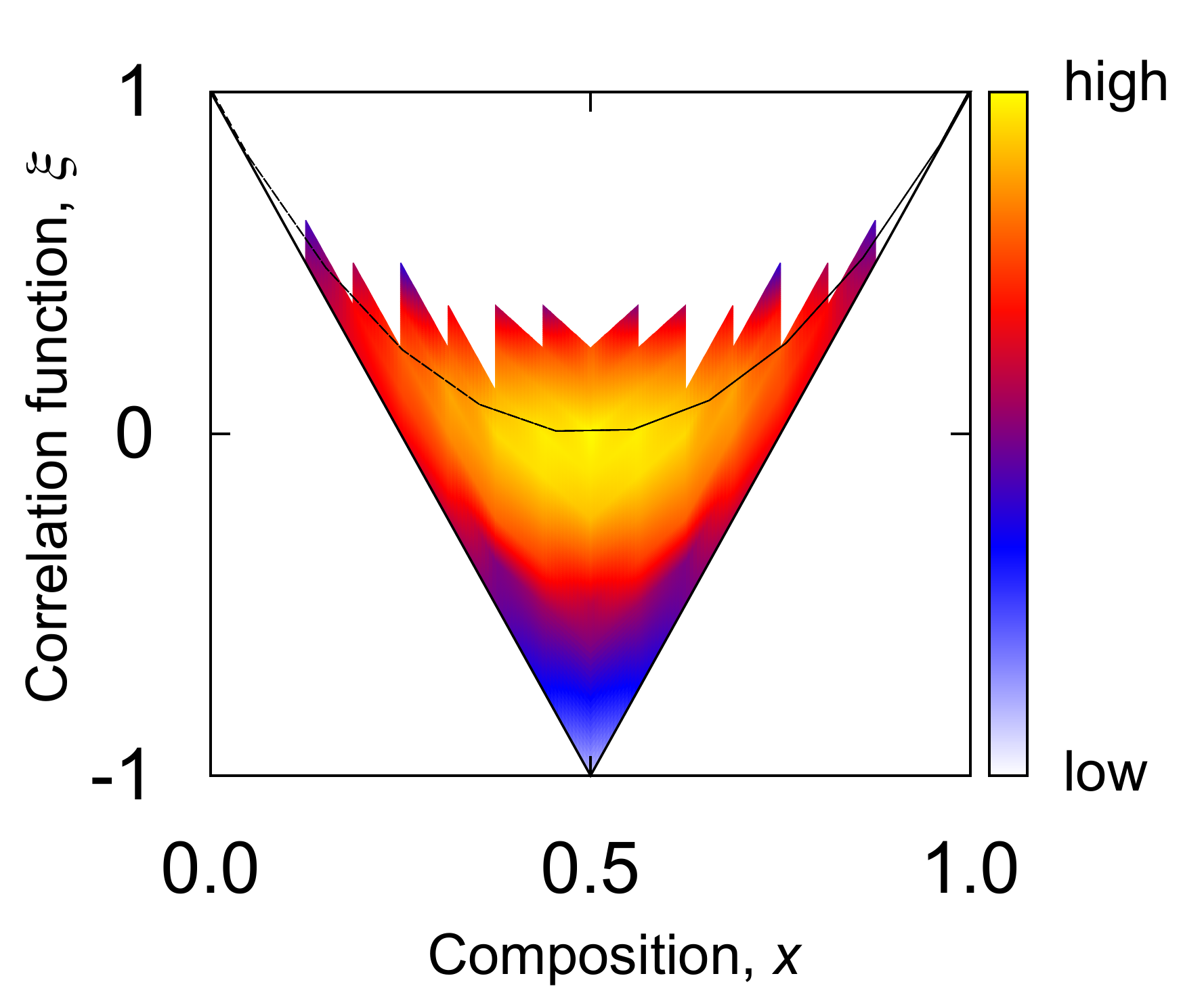}%
	\caption{\label{fig:2ddos}
	The number of all possible states in terms of $x$ and $\xi$ in $4 \times 4$ 2D square lattice Ising model. An open triangle and a quadratic curve denote the expected limiation for $x$ and $\xi$ at thermodynamic limit and $\mu_k$ in Eq.~(\ref{eq:average}).
	}
\end{figure}

In order to caluclate Eq.~(\ref{eq:SI_mod}), intervals of integration should be determined as in terms of $\xi_k$.
In our previous study in CuAu\cite{0953-8984-27-38-385201}, we determined the intervals of integration as $-3\sigma_k < \xi_k < -3\sigma_k$ for any $k$.
This is because P($\vec{\xi}$) obeys to multivariate NDF.
Meanwhile, in this study, we choose a model system, 2D square lattice Ising model that does not have geometrical frustration and whose configurational polyhedron\cite{order-and-phase-stability} can be easily obtained.
In $4 \times 4$ square lattice Ising model, the exact DOS in terms of $x$ and $\xi$ is obtained through counting all possible states, and is shown in Fig.~\ref{fig:2ddos}.
As shown in Fig.~\ref{fig:2ddos}, we can easily estimate the intervals of integration at thermodynamic limit.
The upper limit of $\xi$ increase toward to $+1$ with the increase of $N$, and the lower limit of $\xi$ is estimated as follows:
\[
  \xi = \begin{cases}
    1-4x & (0\leq x \leq0.5) \\
    4x-3 & (0.5<x \leq 1)
  \end{cases}
\]
We show these limitation of $\xi$s in Fig.~\ref{fig:2ddos} as straight black lines, and also show Eq.~(\ref{eq:average}) as a quadratic curve.
These limitations of $\xi$s are chosen in this study.

Since free energy estimated by SI does not depend on system size, for practical use, we can employ SI for any finite system size, which should always give identical result at thermodynamic limit.
However, suppose the fact that the DOS in terms of $\xi$ numerically obey to NDF is valid at thermodynamic limit, it is natural to employ SI in large system size.
Using SI, We obtained free energies in the $L \times L$ Ising square lattice for each $L = 16,32,64$ and $128$ with nearest neighbor coupling constants $J=\pm0.01$ where $J>0$ and $J<0$ correspond to antiferro and ferro magnetism, and confirmed the converged result in $L = 128$.

\begin{figure}
	\centering
	\includegraphics[width=0.8\columnwidth]{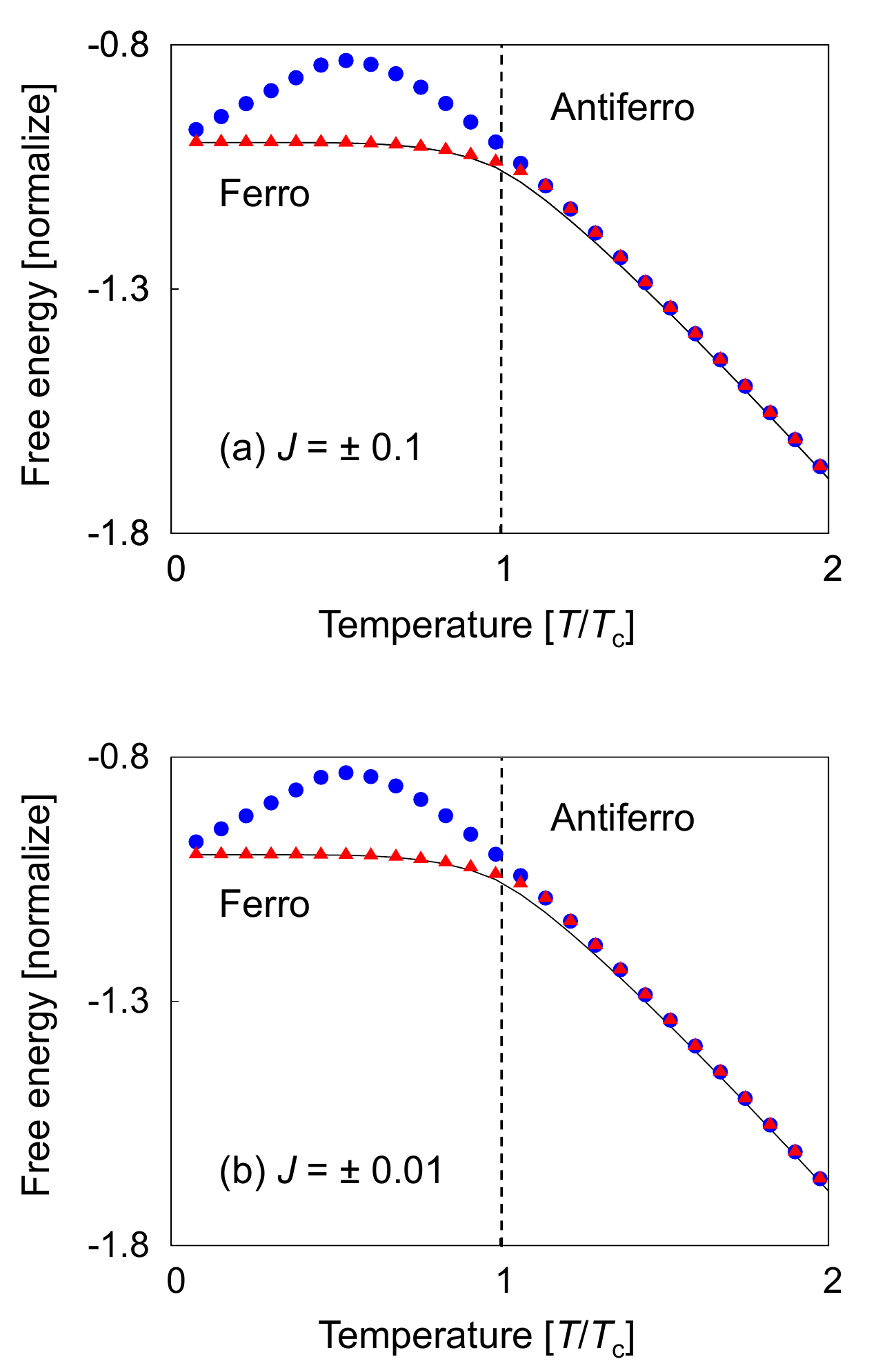}%
	\caption{\label{fig:result}%
	Free energy 2D square lattice Ising model where (a) $J=\pm 0.1$ and (b) $J=\pm 0.01$. Solid line and points denote exact solution and the result using out method in $ 128 \times 128 $ square lattice for ferro and anti ferro magnetism.
	}%
\end{figure}

In Fig.~\ref{fig:result}, free energy via exact and our method are shown where (a) $J=\pm0.1$ and (b) $J=\pm0.01$.
Here we use a normalized unit that $F$ is divided by $J$ because
Fig.~\ref{fig:result} clearly shows that changes in $J$ make no difference to the normalized $F$.
Note that exact result is symmetric for $J$ but our results are not.
This is because the DOS in terms of $x$ and $\xi$ is symmetric for $\mu$, and $\mu$ is absolute value when the number atoms in cluster $k$ is even (see Eq.~(\ref{eq:average})).
Fig.~\ref{fig:result} shows that above $T_{\rm c}$, SI evaluate magnetic free energy accurately for both ferro and anti-ferro magnetism.
This accurate prediction of $F$ above $T_{\rm c}$ has the same tendency for alloy system with constant composition, which has been confined by our previous study\cite{0953-8984-27-38-385201}.
In particular for ferro magnetism system, SI successfully reproduces the free energy not only above $T_{\rm c}$ but also under $T_{\rm c}$.
This means that our {\it a priori} known DOS efficiently describe the tendency that all spins are up or down and disorder.

\section{\label{sec:level4} summary}
% Improved method is proposed to evaluate configurational free energy directly for large systems on given lattice in any composition.
% The present method express the density of states(DOS) as the parameters independent of system but depend on lattice, and replace the DOS unknown a priori as that known a priori.
% Through two-dimensional square lattice Ising model, we find that the present method can evaluate configurational free energy accurately .
% This result also implies a new landscape describing the density of states by the distribution function known a priori in phase space restricted on given lattice.

We propose an extended method to estimate configurational free energy directly for any composition, lattice, and the number of interactions.
% The new expression of $Z$ in Eq.~(\ref{eq:SI_mod}), $Z$ is separated into the variables that depend on constituent elements and are independent of those.
The new expression of $Z$ in Eq.~(\ref{eq:SI_mod}) that known {\it a priori} is valid for 2D square lattice Ising model above critical temperature.

\begin{acknowledgments}
This work was supported by a Grant-in-Aid for Scientific Research on Innovative Areas “Materials Science on Synchronized LPSO Structure” (26109710) and a Grant-in-Aid for Young Scientists B (25820323) from the MEXT of Japan, Research Grant from Hitachi Metals$\cdot$Materials Science Foundation, and Advanced Low Carbon Technology Research and Development Program of the Japan Science and Technology Agency (JST).
\end{acknowledgments}

\bibliography{paper.bib}

\end{document}